\documentclass[cameraready]{Interspeech}
\usepackage{booktabs, multirow}

\usepackage{siunitx}
\title{Speech Entrainment in Multi-Party Conversations with a Digital Agent}

\author[affiliation={1}, orcid=0009-0005-8127-586X, correspondingauthor]{Nicholas}{Mehlman}
\author[affiliation={1}, orcid=0009-0002-3401-8607]{Kaitlin}{Zareno$^*$}
\author[affiliation={1}, orcid=0000-0003-0308-795X]{Kleanthis}{Avramidis$^*$}
\author[affiliation={1}, orcid=0009-0001-5780-9797]{Anfeng}{Xu}
\author[affiliation={1}, orcid=0000-0002-1052-6204]{Shrikanth}{Narayanan}

\address{
    $^1$ University of Southern California\\
    $^*$ \textit{equal contribution}
}

\email{nmehlman@usc.edu, zareno@usc.edu, avramidi@usc.edu, anfengxu@usc.edu, shri@usc.edu}


\usepackage{comment}

\begin{document}

\maketitle

\begin{abstract}
    It has been widely observed that individuals engaged in conversation tend to adapt their speaking style to more closely match the other interlocutors. However, most prior work has focused on dyadic interactions among humans. In this paper, we investigate entrainment effects in a novel setting: a multi-party interaction between groups of humans and a digital agent. This scenario offers important insights into how the participation of non-human actors modulates both short-time and temporally resolved conversational dynamics. To address these questions, we collect and analyze a unique dataset that consists of both adult and family (parent/children) sessions, enabling us to examine how entrainment manifests differently across these cohorts. We consider a range of knowledge-driven and model-based entrainment features and find that, while individuals locally entrain with other humans, global entrainment, and entrainment with the agent remains limited and cohort-dependent.
\end{abstract}

\section{Introduction}
\label{sec:intro}

Entrainment, i.e., the process by which speakers adopt similar speaking styles during a conversation, is a widely observed phenomenon in the sociolinguistic literature~\cite{gallois2005communication}. Researchers have observed that a participant's speech behavior, including pitch and speaking rate, tends to converge with that of the other interlocutors~\cite{WYNN2022101173, levitan2012acoustic}. To date, the majority of entrainment-based analyses have been conducted within the relatively limited context of dyadic human communication. However, recent advancements in the field of voice-based AI agents have introduced non-human participants into our spoken interactions. While current engagement with these agents is predominantly one-on-one (e.g., talking with a smart assistant or an automated help service), it is almost certain that these agents will also permeate more complex contexts, including those involving multiple parties and participants of various ages. 

Understanding whether and how entrainment effects apply in these novel situations will be crucial to understanding how the presence of a digital agent impacts conversation dynamics, knowledge that will help develop agents capable of effective communication with humans across a wide range of scenarios. To address these challenges, this work considers the largely unexplored intersection between multiparty conversational interactions and conversations with a digital agent. We first collect a novel dataset of multiparty interactions with a digital agent, both with groups of adults and also families (parents and children). We then apply feature-based as well as deep learning-based methods to analyze acoustic and semantic entrainment within human participants and between humans and the agent.

\begin{table}[t]
\centering
\footnotesize
\caption{Summary dataset statistics.}
\label{tab:session_stats}
\begin{tabular}{lcc}
\toprule
\textbf{Summary metric} & \textbf{Adults} & \textbf{Families} \\
\midrule
Number of sessions & 30 & 10 \\
Median turns per session & 24.5 & 28 \\
Average turn duration (s) & 26.03 & 18.05 \\
Min speakers per session & 2 & 2 \\
Max speakers per session & 6 & 3 \\
\midrule
Total duration (min) & 337.61 & 65.88\\
\bottomrule
\vspace{-1cm}
\end{tabular}
\end{table}

\subsection{Background}

Conversational entrainment has been widely demonstrated in a variety of different linguistic, syntactic, and acoustic aspects such as pitch, rhythm, and even word choice and phonetic pronunciation \cite{WYNN2022101173, levitan2012acoustic, pardo2006phonetic}. Higher degrees of entrainment have been shown to be a marker of effective communication \cite{lee2010quantification,Lee2014ComputingVocalEntrainment} as well as social desirability \cite{Natale1975CONVERGENCEOM, michalsky2017pitch}. While research on entrainment has largely focused on the dyadic setting, several works have considered entrainment in multi-party scenarios. For example, Rahimi et al.~\cite{rahimi2017entrainment} found that teams of participants engaged in a cooperative board game exhibited convergence on both the acoustic and lexical levels, and further demonstrated that the degree of entrainment was predictive of their team's effectiveness. Many methods of measuring entrainment, such as those used in \cite{levitan2012acoustic,Levitan2011MeasuringAE}, leverage hand-crafted audio features, such as pitch, amplitude, jitter, and speaking rate. The use of deep learning features for entrainment analysis has been previously explored in works such as \cite{nasir2018towards}  and \cite{nasir2022}, which used unsupervised learning approaches to learn an embedding space that captured entrainment distances between speakers. 

Several recent efforts have also considered the role of entrainment in conversation between humans and digital agents. For example, \cite{Gessinger22} ran a series of experiments with a digital language tutor that found that subjects did tend to adapt their speaking to match that of the agent. Turn-level entrainment with Amazon Alexa was also observed in \cite{alexa-speech-rate}, which also reported general reductions and speaking rate and greater hyperarticulation compared to their baseline. A larger body of work (e.g., \cite{levitan2016implementing, 10.11450}) has tackled the problem of designing digital agents that engage with users as a means to improve conversational quality and user trust. However, works such as \cite{levitan2016implementing} and \cite{benus2018prosodic}, have found that the impact of agent-to-human entrainment is complex and not universally beneficial. For example, \cite{benus2018prosodic} observed that female users actually preferred dis-entraining avatars over entraining ones. Specific work on children is limited, but \cite{GAMPE2023107693} found that children do modify their speech when talking to a digital agent instead of human partners, and \cite{Coulston2002AmplitudeCI} similarly demonstrated child-agent convergence in speech amplitude.

\section{Methods}
\subsection{Data Collection}

For our experiment, we recruited 2 separate cohorts of English-speaking participants. The first cohort consisted of young adults (largely undergraduate and master's students from a major university), while the second consisted of families with young children between the ages of 8 and 14. Each cohort participated in an 8 to 12-minute multiparty conversation (2 to 6 human speakers)  with an animated digital agent. Different scripts were used for each of the cohorts to ensure that the interactions were both engaging and age-appropriate. However, in both cases, the agent, who took the form of an alien visiting Earth, asked the subjects a series of questions about their daily experience in school/university, while not-so-subtly hinting that it had nefarious plans. Participants responded either individually or, when prompted, after discussing the answer as a group.

As is common practice in a Wizard of Oz setup, each of the agent's responses was selected from a dialog tree by a human operator who monitored the interaction from a separate room. Audio was recorded as a single channel using a beamformed microphone array as well as lavalier mics to assist with annotations. The agent's speech was automatically removed from the captured audio, but was cached directly along with timestamps for easy reconstruction of the conversation turns. 

Our experimental team manually annotated the collected recordings to assign timestamps and speaker labels to each utterance. Using these annotations, we segmented the single-channel array recording into different files that isolated the speech from each of the (human) speakers. Periods during which a given speaker was inactive were replaced with silence. The agent's audio was automatically saved by the recording system. To further enable granular data analysis, we partitioned each of these recordings into conversational turns, which consisted of one agent prompt (typically a question) followed by the participants' responses. Table \ref{tab:session_stats} summarizes the details of the final dataset for both the adult and family cohorts.

\subsection{Entrainment Analysis}

We considered two distinct hypotheses regarding the groups' conversational entrainment, based on two different interaction timescales:
\begin{itemize}
    \item \textbf{H1: Local Entrainment:\,} participants entrain to the local conversational and linguistic style of the current turn.
    \item  \textbf{H2: Global Entrainment:\,} participants exhibit greater degrees of entrainment as the session progresses. 
\end{itemize}
We tested those hypotheses across several different conversational cohorts. For both the adult and family sessions, we consider both participant-to-participant (P2P) entrainment and participant-to-agent (P2A) entrainment. We additionally consider the child-to-guardian (C2G) and child-to-agent (C2A) sub-groups within the family sessions. 

We extracted a variety of features for entrainment analysis, as prior work such as \cite{OSTRAND2021101074} has indicated that entrainment effects often occur in only a subset of speech attributes.

\subsubsection{Handcrafted Feature Extraction}

\textbf{Amplitude\,}
We computed frame-level RMS amplitude for each participant and the agent for each conversational turn. From these amplitude values, we computed five statistical features (mean, standard deviation, min, max, and range) per turn and speaker. Non-speech segments were discarded.

\noindent\textbf{Pitch\,}
We used PESTO~\cite{riou2025pesto} to estimate the frame-level pitch (F$0$) for both the participant and agent channels. Out of these, we computed turn-level statistical features (mean, standard deviation, min, max, and range), again excluding silence.

\noindent\textbf{Emotion\,}
We used VoxProfile \cite{feng2025voxprofilespeechfoundationmodel} to compute turn-level arousal, valence, and dominance values. More specifically, we used both the Whisper-based and WavLM-based models, averaging their predictions to produce the final feature.
 
\subsubsection{Speech Foundation Models}

\textbf{Emotion\,}
We extracted embeddings from the penultimate layer of the VoxProfile emotion models. The embeddings from the Whisper and WavLM versions were concatenated to generate an aggregate embedding. 

\noindent\textbf{Whisper~\cite{whisper}\,} We extracted embeddings from the encoder of \textit{Whisper-base}. To obtain a fixed-length representation per turn, output activations were averaged across time. We expect these representations to primarily capture semantic attributes.

\noindent\textbf{Mimi~\cite{defossez2024moshi}\,} Mimi embeddings were obtained using the neural-codec encoder from the \textit{Moshi} checkpoint in HuggingFace\footnote{kyutai/moshiko-pytorch-bf16}. The speech waveforms were resampled to 24 kHz and passed through the Mimi feature extractor to obtain latent embeddings. To obtain a fixed-length representation per turn, output activations were averaged across time. These representations were intended primarily to capture the phonetic aspects of the speech.

\section{Local Entrainment}

\subsection{Methodology} \label{sec:local-methods}

We computed pairwise differences between features for a given interaction mode (e.g., P2P vs. P2A) both within a given turn and across different turns within the same session. To assess local entrainment, we tested whether features were more similar (i.e., smaller differences) in the within-turn case. This was done using a mixed-effects regression model with a random intercept for each session/speaker-pair combination. To make this explicit, consider utterance $i$ from speaker $a$ and utterance $j$ from speaker $b$, both in session $s$. We model the difference $\Delta_{s,i,j}^{(f)}=\mathrm{dist}(f_{i,s} - f_{j,s})$ in feature $f$ as
$$\Delta_{s,i,j}^{(f)}=\beta_{s,a,b} + \gamma \mathrm{I}(i,j)+\epsilon$$
where $\beta_{s,a,b}$ is the intercept for this speaker pairing, $\mathrm{I}(i,j)$ indicates whether or not utterances $i$ and $j$ came from the same turn, and $\epsilon$ is a residual noise term. We compute both the value and statistical significance of $\gamma_t$, which indicates how the within vs. cross-turn scenarios impacted feature differences. Bonferroni correction~\cite{bonferroni1936teoria} was applied within each feature category to correct for multiple tests on different features. Here and in the following, we do not normalize features, and hence $\gamma$ values reflect the original scale of each feature.

\begin{table*}[t]
\caption{\textbf{Local} entrainment effects for hand-crafted features in the adult and family sessions. The coefficient $\gamma$ reflects the within-turn fixed effect, i.e., $\gamma <$ 0 indicates that the within-turn differences were smaller than the cross-turn ones. * denotes $p<0.0001$ after Bonferroni correction, while \textbf{bold} values were significant. P2P refers to all human speaker pairs, whereas C2G refers to all pairs including one child and one guardian.}
\label{tab:main-local-acu}
\centering
\small
\setlength{\tabcolsep}{6pt}
\renewcommand{\arraystretch}{1.15}

\begin{tabular}{
ll
S[table-format=-2.4] S[table-format=1.4]
S[table-format=-2.4] S[table-format=1.4]
S[table-format=-2.4] S[table-format=1.4]
}
\toprule
& & \multicolumn{2}{c}{\textbf{Adults}} 
  & \multicolumn{4}{c}{\textbf{Families}} \\
\cmidrule(lr){3-4}\cmidrule(l){5-8}
& & \multicolumn{2}{c}{P2P}
  & \multicolumn{2}{c}{P2P}
  & \multicolumn{2}{c}{C2G} \\
\cmidrule(lr){3-4}\cmidrule(lr){5-6}\cmidrule(lr){7-8}
\textbf{Category} & \textbf{Feature}
& {$\gamma$} & {p-value}
& {$\gamma$} & {p-value}
& {$\gamma$} & {p-value} \\
\midrule

\multirow{3}{*}{Amplitude}
& mean  & -0.0091 & {\textbf{*}}
        & -0.0042 & 0.4267
        & -0.0044 & 0.5008 \\
& st. deviation   & -0.0087 & {\textbf{*}}
        & -0.0012 & 1.0000
        & -0.0012 & 1.0000 \\
& range & -0.0346 & {\textbf{*}}
        & -0.0080 & 1.0000
        & -0.0050 & 1.0000 \\

\midrule

\multirow{3}{*}{Pitch~\cite{riou2025pesto}}
& mean  & -25.7837 & {\textbf{*}}
        & -10.8488 & 0.0814
        & -12.8029 & 0.0505 \\
& st. deviation   & -23.7777 & {\textbf{*}}
        & -5.4494 & 1.0000
        & -5.7072 & 1.0000 \\
& range & -81.7369 & \textbf{0.0047}
        & -47.2254 & 1.0000
        & -46.1026 & 1.0000 \\

\midrule

\multirow{3}{*}{Emotion~\cite{feng2025voxprofilespeechfoundationmodel}}
& arousal  & -0.0388 & {\textbf{*}}
            & -0.0274 & \textbf{*}
            & -0.0303 & \textbf{*} \\
& valence  & -0.0276 & {\textbf{*}}
            & -0.0096 & 0.0853
            & -0.0088 & 0.2332 \\
& dominance & -0.0284 & {\textbf{*}}
            & -0.0196 & \textbf{*}
            & -0.0211 & \textbf{*} \\

\bottomrule
\end{tabular}
\end{table*}

\begin{table*}[t]
\caption{Deep learning-based \textbf{local} entrainment effects for adult and family sessions. The coefficient $\gamma$ reflects the within-turn fixed effect, i.e., $\gamma <0$ indicates that the within-turn differences were smaller than the cross-turn ones. * denotes $p<0.0001$ after Bonferroni correction, while \textbf{bold} values were significant. P2P refers to all human speaker pairs, whereas C2G refers to all pairs including one child and one guardian. }
\label{tab:main-local-dl}
\centering
\small
\setlength{\tabcolsep}{6pt}
\renewcommand{\arraystretch}{1.15}

\begin{tabular}{
ll
S[table-format=-2.4] S[table-format=1.4]
S[table-format=-2.4] S[table-format=1.4]
S[table-format=-2.4] S[table-format=1.4]
}
\toprule
& & \multicolumn{2}{c}{\textbf{Adults}} 
  & \multicolumn{4}{c}{\textbf{Families}} \\
\cmidrule(lr){3-4}\cmidrule(l){5-8}
& & \multicolumn{2}{c}{P2P}
  & \multicolumn{2}{c}{P2P}
  & \multicolumn{2}{c}{C2G} \\
\cmidrule(lr){3-4}\cmidrule(lr){5-6}\cmidrule(lr){7-8}
\textbf{Category} & \textbf{Feature}
& {$\gamma$} & {p-value}
& {$\gamma$} & {p-value}
& {$\gamma$} & {p-value} \\
\midrule

\multirow{2}{*}{Emotion~\cite{feng2025voxprofilespeechfoundationmodel}}
& dist$_{\cos}$ & -0.1089 & {\textbf{*}}
            & -0.0546 & \textbf{0.0019}
            & -0.0557 & \textbf{0.0055} \\
& dist$_{\ell_2}$ & -0.8132 & {\textbf{*}}
            & -0.4645 & \textbf{0.0005}
            & -0.4557 & \textbf{0.0030} \\

\midrule

\multirow{2}{*}{Whisper~\cite{whisper}}
& dist$_{\cos}$ &-0.013418 & \textbf{*}
            & -0.010741 & * 
            & -0.011007 & * \\
& dist$_{\ell_2}$ & -1.371361 & \textbf{*}
            & -1.118139	 & *
            & -1.156685 & * \\

\midrule

\multirow{2}{*}{Mimi~\cite{defossez2024moshi}}
& dist$_{\cos}$ & -0.028699 & \textbf{*}
            & -0.042071 & *
            & -0.041729& * \\
& dist$_{\ell_2}$ & -0.108165 & \textbf{*}
            & -0.153524	 & *
            & -0.152665 & * \\

\bottomrule
\vspace{-1mm}
\end{tabular}
\end{table*}

\subsection{Results}

Tables \ref{tab:main-local-acu} and \ref{tab:main-local-dl} report the measured local entrainment effect $\gamma$ along with statistical significance for the handcrafted and deep learning features, respectively. A negative value of $\gamma$ indicates that the within-turn differences were smaller than the cross-turn ones. Note that we drop the P2A and C2A interaction modes for which no significant entrainment effects were observed. The adults exhibit strong local entrainment, with almost all measured features showing a statistically significant effect. This suggests that the adult participants reliably adapted various aspects of their speaking style (pitch, emotion, etc.) to align with the other human speakers in a given turn. Given that the participants were largely strangers prior the the session, it may be that this response arises from a desire to increase social desirability \cite{Natale1975CONVERGENCEOM,michalsky2017pitch}, and improve group cohesion \cite{rahimi2017entrainment}. This is supported by the results of our post-interaction survey, in which $70\%$ reported agreeing or strongly agreeing that their group had a good rapport, and $62\%$ reported agreeing or strongly agreeing that their group was in sync with one another.

Unlike the adult sessions, the family sessions exhibit no amplitude or pitch entrainment; however, we do find evidence for entrainment of emotional features (except valence), as well as deep-learning embeddings. These results apply to both the P2P and C2G interaction modes. This finding may suggest that the children are less attuned to granular vocal attributes, such as pitch and amplitude, and instead track the higher-level (and more salient) emotional cues of their fellow participants. However, the C2G entrainment does imply that the children were responsive to emotional and verbal cues from their caregiver.

Finally, it is worth noting that neither group (adults or families) showed local entrainment with the agent, implying that they instinctively treated the agent differently from other human partners. This may be in part due to the structure of the interaction, in which the agent asked various questions of the group, and thereby establishes unique conversational niches for the humans and the agent, respectively. However, it is also plausible that participants, consciously or unconsciously, recognize the agent as non-human and hence do not feel compelled to engage in the same rapport-building behavior.  

\section{Global Entrainment}

\subsection{Methodology}

We computed inter-turn pairwise differences between features as in \ref{sec:local-methods}. However, rather than comparing them to cross-turn differences, we instead compare the first $5$ turns to the last $5$ turns. We use a mixed-effects regression model with a random intercept for each session/speaker-pair combination, with the fixed effect indicating whether the turn was in the final $5$ rather than the first $5$. As before, Bonferroni correction was applied to correct for multiple tests on different features.

\subsection{Results}

\begin{table}[t]
\caption{Participant-to-participant (P2P) \textbf{global} entrainment for adults and families sessions. The coefficient $\gamma$ reflects the fixed effect of the last $5$ turns, i.e., $\gamma<0$ indicates that feature differences were smaller during the last 5 turns. Omitted feature categories were found to be overall not significantly affected. \textbf{Bold} values were significant after Bonferroni correction.}
\label{tab:global-p2p}
\centering
\small
\setlength{\tabcolsep}{3.6pt}
\renewcommand{\arraystretch}{1.15}
\begin{tabular}{llcccc}
\toprule
\textbf{Category} & \textbf{Feature}
& \multicolumn{2}{c}{\textbf{Adults P2P}}
& \multicolumn{2}{c}{\textbf{Families P2P}} \\
\cmidrule(lr){3-4}\cmidrule(l){5-6}
& & ${\gamma}$ & {p-value} & ${\gamma}$ & {p-value} \\
\midrule

\multirow{1}{*}{Pitch}
& mean  & -36.1776 & \textbf{0.0409}  & 14.5254  & 1.0000 \\

\midrule

\multirow{2}{*}{Emotion}
& dist$_{\ell_2}$  & 0.2870 & 0.2797 & 0.5813 & 0.5647 \\
& dist$_{\cos}$ & 0.05076 & \textbf{0.0418} & 0.0861 & 0.0636 \\

\midrule

\multirow{2}{*}{Whisper}
& dist$_{\ell_2}$  & 0.3819 &
 0.1005  &  0.4124 &  0.3831 \\
& dist$_{\cos}$   & 0.0039 &
 0.1124 &  0.0052 & 0.1533
  \\

\midrule

\multirow{2}{*}{Mimi}
& dist$_{\ell_2}$  & -0.0425 & 0.0521
& 0.0414 &  0.3389 \\
& dist$_{\cos}$   & -0.0095 &  \textbf{0.0412} & 0.0036 & 0.6584 \\

\bottomrule
\end{tabular}
\end{table}

Table \ref{tab:global-p2p} reports the coefficients and p-values for global entrainment between participants (P2P). As before, a negative value of $\gamma$ indicates that feature differences were smaller during the last $5$ turns, reflecting entrainment. Note that we drop hand-crafted features for which none of the effects were significant to aid readability. Interestingly, the adult participants show only weak global entrainment effects in their mean pitch/f0 and the Mimi phonetic embeddings. The disentrainment ($\gamma>0$) for the emotion embeddings may be a manifestation of differing reactions to the novel experience of conversing with a non-human interlocutor. For example, some individuals may have found the character endearing while others may have found it awkward or even unnerving. The family sessions exhibit no significant global entrainment effects across any of the features tested.


\begin{table}[t]
\caption{\textbf{Global} child-to-agent (C2A) entrainment for families sessions. The coefficient $\gamma$ reflects the fixed effect of the last $5$ turns, i.e., $\gamma<0$ indicates smaller feature differences during the last 5 turns. Omitted feature categories were found to be overall not significantly affected. \textbf{Bold} values were significant after Bonferroni correction.}
\label{tab:global-c2a}
\centering
\renewcommand{\arraystretch}{1.15}
\setlength{\tabcolsep}{4.5pt}
\begin{tabular}{llcc}
\toprule
\textbf{Category} & \textbf{Feature}
& ${\gamma}$ & p-value \\
\midrule
\multirow{3}{*}{Amplitude}
& mean & 0.0181 & \textbf{0.0182} \\
& st. deviation & 0.0152	& \textbf{0.0097} \\
& range & 0.0566	& \textbf{0.0045} \\
\midrule

\multirow{2}{*}{Whisper}
& dist$_{\ell_2}$  & -0.9678 & \textbf{0.0340}
 \\
& dist$_{cos}$  & -0.0189
 & \textbf{0.0330} \\

\midrule

\multirow{2}{*}{Mimi}
& dist$_{\ell_2}$  &  -0.0429  & 0.5247\\
& dist$_{cos}$   &  -0.0272  & 0.3841 \\
\bottomrule
\end{tabular}
\end{table}

Table \ref{tab:global-c2a} shows the observed global entrainment between the children in the family session and the agent (C2A)\footnote{We isolate this interaction mode, as the effects for others were overwhelmingly not significant or else were driven by the C2A grouping.}. The children exhibit significant entrainment in Whisper features, but show disentrainment in amplitude. One possible explanation for the semantic (Whisper) convergence (not observed for the guardians or the adult cohort) is that the children form a stronger emotional bond with the agent, as a result of being more engrossed in the experience. However, for the amplitude features, a regression analysis indicated that the children increased their mean amplitude ($p = 0.001$), amplitude standard deviation ($p = 0.0031$), and amplitude range ($0.0010$) as the interaction progressed. This finding, along with the fact that emotional dominance ($p=0.0225$) also increased, is highly consistent with their becoming increasingly confident and assertive throughout the session. This evolution likely creates the appearance of amplitude decoupling, since the amplitude features of the agent's speech remain stable over time.

\section{Discussion and Conclusions}

While our findings reveal strong inter-group entrainment effects on a local level, global inter-group entrainment was minimal for the adult cohort, and non-existent for the families. This difference may be attributable to the experimental procedure, in particular the fact that the subjects spent approximately 10 minutes together prior to the start of data collection, during which they learned about the study and signed the necessary consent forms. This might have put the group in a state of partial or full entrainment by the time data collection began. The fact that global entrainment was absent for the families supports this hypothesis, since they, unlike the adults, generally traveled to the study together, increasing the time window for pre-collection synchrony. Future studies could address this by separating the participants for some duration of time prior to the experiment.

Overall, inter-group dynamics are similar for both the adult and family sessions, although locally the children appear to respond to more salient speech cues such as emotion. Furthermore, only the children exhibit significant entrainment with the agent, which may imply that children are somewhat more prone to see the agent as an equal participant in the conversation \cite{benus2014entrainment}. This effect, however, is only observed under a long-term (global) analysis which seems to indicate that they also require time to become comfortable with the novel scenario.

The present findings should be interpreted in light of several limitations. First, the number of sessions—particularly in the family cohort—is relatively modest, which may limit statistical power for detecting global entrainment effects. Second, the interaction paradigm was highly structured, with the agent consistently occupying an interviewing role, which might limit generalization to more spontaneous multiparty conversations. Finally, while foundation-model embeddings provide rich representational similarity measures, their dimensions are not directly interpretable, and likely inter-mingle different speech attributes that might correspond to the reported entrainment effects, aside from their assigned role (i.e., semantic for Whisper embeddings). Overall, future work should explore similar effects in more open-ended contexts and modulating factors.

Overall, these findings highlight the importance of social context, participant age, and interaction structure in shaping adaptation dynamics, and underscore the need for further work examining how digital agents can be designed to more naturally integrate into complex, multi-party human conversations.

\newpage


\section{Acknowledgments}

Thanks to Disney Research for their help in the study design and data collection process. 

\section{Generative AI Use Disclosure}

Generative AI tools were used in this study to assist with language polishing, manuscript editing, and assisting in code implementations to perform analyses and visualizations relevant to this paper. These tools were not prompted for results generation, data analysis, data interpretation, or at any stage of data collection. All authors are fully aware of the extent of generative AI use in this work, take full responsibility for the content of the manuscript, and consent to its submission to Interspeech.

\bibliographystyle{ieeetr}
\bibliography{refs}

@article{Lee2014ComputingVocalEntrainment,
 author = {Lee, Chi-Chun and Katsamanis, Athanasios and Black, Matthew P. and Baucom, Brian and Christensen, Andrew and Georgiou, Panayiotis and Narayanan, Shrikanth S.},
 bib2html_rescat = {bsp},
 doi = {10.1016/j.csl.2012.06.006},
 journal = {Computer, Speech, and Language},
 link = {http://sail.usc.edu/publications/files/JeremyLee-CSL2014-ProsodyEntrainment.pdf},
 month = {mar},
 number = {2},
 pages = {518-539},
 title = {Computing Vocal Entrainment: A Signal-derived PCA-based Quantification Scheme with Application to Affect Analysis in Married Couple Interactions},
 volume = {28},
 year = {2014}
}

@inproceedings{Coulston2002AmplitudeCI,
  title={Amplitude convergence in children's conversational speech with animated personas},
  author={Rachel Coulston and Sharon L. Oviatt and Courtney Darves},
  booktitle={Interspeech},
  year={2002},
  url={https://api.semanticscholar.org/CorpusID:392412}
}

@inproceedings{michalsky2017pitch,
  title={Pitch convergence as an effect of perceived attractiveness and likability.},
  author={Michalsky, Jan and Schoormann, Heike},
  booktitle={Interspeech},
  pages={2253--2256},
  year={2017}
}

@inproceedings{whisper,
  title={Robust speech recognition via large-scale weak supervision},
  author={Radford, Alec and Kim, Jong Wook and Xu, Tao and Brockman, Greg and McLeavey, Christine and Sutskever, Ilya},
  booktitle={International Conference on Machine Learning},
  pages={28492--28518},
  year={2023},
  organization={PMLR}
}

@article{pardo2006phonetic,
  title={On phonetic convergence during conversational interaction},
  author={Pardo, Jennifer S},
  journal={The Journal of the acoustical society of America},
  volume={119},
  number={4},
  pages={2382--2393},
  year={2006},
  publisher={Acoustical Society of America}
}

@inproceedings{nasir2018towards,
  title={Towards an Unsupervised Entrainment Distance in Conversational Speech Using Deep Neural Networks},
  author={Nasir, Md and Baucom, Brian and Narayanan, Shrikanth and Georgiou, Panayiotis},
  booktitle={Proc. Interspeech 2018},
  pages={3423--3427},
  year={2018}
}

@inproceedings{Levitan2011MeasuringAE,
  title={Measuring Acoustic-Prosodic Entrainment with Respect to Multiple Levels and Dimensions},
  author={Rivka Levitan and Julia Hirschberg},
  booktitle={Interspeech},
  year={2011},
  url={https://api.semanticscholar.org/CorpusID:14147636}
}

@ARTICLE{nasir2022,
  author={Nasir, Md and Baucom, Brian and Bryan, Craig and Narayanan, Shrikanth and Georgiou, Panayiotis},
  journal={IEEE Transactions on Affective Computing}, 
  title={Modeling Vocal Entrainment in Conversational Speech Using Deep Unsupervised Learning}, 
  year={2022},
  volume={13},
  number={3},
  pages={1651-1663},
  doi={10.1109/TAFFC.2020.3024972}}

@ARTICLE{alexa-speech-rate,
  
AUTHOR={Cohn, Michelle  and Liang, Kai-Hui  and Sarian, Melina  and Zellou, Georgia  and Yu, Zhou },
         
TITLE={Speech Rate Adjustments in Conversations With an Amazon Alexa Socialbot},
        
JOURNAL={Frontiers in Communication},
        
VOLUME={Volume 6 - 2021},

YEAR={2021},

URL={https://www.frontiersin.org/journals/communication/articles/10.3389/fcomm.2021.671429},

DOI={10.3389/fcomm.2021.671429},

ISSN={2297-900X},

}

@article{GAMPE2023107693,
title = {How children speak with their voice assistant Sila depends on what they think about her},
journal = {Computers in Human Behavior},
volume = {143},
pages = {107693},
year = {2023},
issn = {0747-5632},
doi = {https://doi.org/10.1016/j.chb.2023.107693},
url = {https://www.sciencedirect.com/science/article/pii/S0747563223000444},
author = {Anja Gampe and Katharina Zahner-Ritter and Joanna Joys Müller and Sarah Rebecca Schmid},
}

@inproceedings{levitan2016implementing,
  title={Implementing Acoustic-Prosodic Entrainment in a Conversational Avatar.},
  author={Levitan, Rivka and Benus, Stefan and G{\'a}lvez, Ramiro H and Gravano, Agust{\'\i}n and Savoretti, Florencia and Trnka, Marian and Weise, Andreas and Hirschberg, Julia},
  booktitle={Interspeech},
  volume={16},
  pages={1166--1170},
  year={2016},
  organization={San Francisco, CA}
}

@phdthesis{Gessinger22,
  author={Iona Gessinger},
  title={Phonetic accommodation of human interlocutors in the context of human-computer interaction},
  year={2022},
  cdate={1640995200000},
  url={https://publikationen.sulb.uni-saarland.de/handle/20.500.11880/32213},
  school={Saarland University, Saarbrücken, Germany}
}

@inproceedings{benus2018prosodic,
  author    = {Be{\v{n}}u{\v{s}}, {\v{S}}tefan and Trnka, Marian and Kuric, Eduard and Mart{\'a}k, Luk{\'a}{\v{s}} and Gravano, Agust{\'i}n and Hirschberg, Julia and Levitan, Rivka},
  title     = {Prosodic entrainment and trust in human-computer interaction},
  booktitle = {Proceedings of Speech Prosody 2018},
  year      = {2018},
  doi       = {10.21437/SpeechProsody.2018-76}
}

@inproceedings{rahimi2017entrainment,
  author    = {Rahimi, Zohreh and Kumar, Anuj and Litman, Diane and Paletz, Susannah and Yu, Mo},
  title     = {Entrainment in Multi-Party Spoken Dialogues at Multiple Linguistic Levels},
  booktitle = {Proceedings of Interspeech 2017},
  pages     = {1696--1700},
  year      = {2017},
  doi       = {10.21437/Interspeech.2017-1568}
}

@article{benus2014entrainment,
  author  = {Be{\v{n}}u{\v{s}}, {\v{S}}tefan},
  title   = {Social Aspects of Entrainment in Spoken Interaction},
  journal = {Cognitive Computation},
  year    = {2014},
  volume  = {6},
  pages   = {802--813},
  doi     = {10.1007/s12559-014-9261-4}
}

@article{Natale1975CONVERGENCEOM,
  title={CONVERGENCE OF MEAN VOCAL INTENSITY IN DYADIC COMMUNICATION AS A FUNCTION OF SOCIAL DESIRABILITY},
  author={Michael Natale},
  journal={Journal of Personality and Social Psychology},
  year={1975},
  volume={32},
  pages={790-804},
  url={https://api.semanticscholar.org/CorpusID:51847153}
}

@inproceedings{lee2010quantification,
  title={Quantification of prosodic entrainment in affective spontaneous spoken interactions of married couples},
  author={Lee, Chi-Chun and Black, Matthew and Katsamanis, Athanasios and Lammert, Adam C and Baucom, Brian R and Christensen, Andrew and Georgiou, Panayiotis G and Narayanan, Shrikanth S},
  booktitle={Proc. Interspeech 2010},
  pages={793--796},
  year={2010}
}

@article{OSTRAND2021101074,
title = {It’s alignment all the way down, but not all the way up: Speakers align on some features but not others within a dialogue},
journal = {Journal of Phonetics},
volume = {88},
pages = {101074},
year = {2021},
issn = {0095-4470},
doi = {https://doi.org/10.1016/j.wocn.2021.101074},
url = {https://www.sciencedirect.com/science/article/pii/S0095447021000462},
author = {Rachel Ostrand and Eleanor Chodroff},
}

@inproceedings{10.11450,
author = {Inden, Benjamin and Malisz, Zofia and Wagner, Petra and Wachsmuth, Ipke},
title = {Timing and entrainment of multimodal backchanneling behavior for an embodied conversational agent},
year = {2013},
isbn = {9781450321297},
publisher = {Association for Computing Machinery},
address = {New York, NY, USA},
url = {https://doi.org/10.1145/2522848.2522890},
doi = {10.1145/2522848.2522890},
booktitle = {Proceedings of the 15th ACM on International Conference on Multimodal Interaction},
pages = {181–188},
numpages = {8},
location = {Sydney, Australia},
series = {ICMI '13}
}

@book{bonferroni1936teoria,
  title={Teoria statistica delle classi e calcolo delle probabilit{\`a}},
  author={Bonferroni, C.E.},
  series={Pubblicazioni del R. Istituto superiore di scienze economiche e commerciali di Firenze},
  url={https://books.google.com/books?id=3CY-HQAACAAJ},
  year={1936},
  publisher={Seeber}
}

@misc{feng2025voxprofilespeechfoundationmodel,
      title={Vox-Profile: A Speech Foundation Model Benchmark for Characterizing Diverse Speaker and Speech Traits}, 
      author={Tiantian Feng and Jihwan Lee and Anfeng Xu and Yoonjeong Lee and Thanathai Lertpetchpun and Xuan Shi and Helin Wang and Thomas Thebaud and Laureano Moro-Velazquez and Dani Byrd and Najim Dehak and Shrikanth Narayanan},
      year={2025},
      eprint={2505.14648},
      archivePrefix={arXiv},
      primaryClass={cs.SD},
      url={https://arxiv.org/abs/2505.14648}, 
}

@article{WYNN2022101173,
title = {Classifying conversational entrainment of speech behavior: An expanded framework and review},
journal = {Journal of Phonetics},
volume = {94},
pages = {101173},
year = {2022},
issn = {0095-4470},
doi = {https://doi.org/10.1016/j.wocn.2022.101173},
url = {https://www.sciencedirect.com/science/article/pii/S0095447022000481},
author = {Camille J. Wynn and Stephanie A. Borrie},
}

@inproceedings{levitan2012acoustic,
  title={Acoustic-prosodic entrainment and social behavior},
  author={Levitan, Rivka and Gravano, Agust{\'\i}n and Willson, Laura and Be{\v{n}}u{\v{s}}, {\v{S}}tefan and Hirschberg, Julia and Nenkova, Ani},
  booktitle={Proceedings of the 2012 Conference of the North American Chapter of the Association for Computational Linguistics: Human language technologies},
  pages={11--19},
  year={2012}
}

@article{gallois2005communication,
  title={Communication accommodation theory},
  author={Gallois, Cindy and Ogay, Tania and Giles, Howard},
  journal={Theorizing about intercultural communication},
  pages={121--148},
  year={2005}
}

@article{riou2025pesto,
  title={PESTO: Real-Time Pitch Estimation with Self-supervised Transposition-equivariant Objective},
  author={Riou, Alain and Torres, Bernardo and Hayes, Ben and Lattner, Stefan and Hadjeres, Ga{\"e}tan and Richard, Ga{\"e}l and Peeters, Geoffroy},
  journal={arXiv preprint arXiv:2508.01488},
  year={2025}
}

@article{defossez2024moshi,
  title={Moshi: a speech-text foundation model for real-time dialogue},
  author={D{\'e}fossez, Alexandre and Mazar{\'e}, Laurent and Orsini, Manu and Royer, Am{\'e}lie and P{\'e}rez, Patrick and J{\'e}gou, Herv{\'e} and Grave, Edouard and Zeghidour, Neil},
  journal={arXiv preprint arXiv:2410.00037},
  year={2024}
}

\end{document}